# Graphene Enhanced Resonant Raman Spectroscopy of Gallium Nitride Nanocrystals

*Marek Kostka[1], Jindřich Mach[1,2]\*, Miroslav Bartošík[1,2,3], David Nezval[2], Martin Konečný[1,2], Vojtěch Mikerásek[1], Linda Supalová[2], Jakub Piastek[1,2] and Tomáš Šikola[1,2]*

[1] Institute of Physical Engineering, Brno University of Technology, Technická 2896/2, 616 69 Brno, Czech Republic

[2] CEITEC BUT, Brno University of Technology, Purkyňova 123, 612 00 Brno, Czech Republic

[3] Department of Physics and Materials Engineering, Faculty of Technology, Tomas Bata University in Zlín, 760 01, Czech Republic

\* Corresponding author: mach@fme.vutbr.cz

**ABSTRACT**

*The scattering of lattice excitations (phonons) with the photoexcited charge carriers is of a major concern in optoelectronic devices. Here, the resonant Raman scattering will be utilized to study an exciton-phonon interaction in GaN nanocrystals, further enhanced by the underlying graphene. Raman spectroscopy using various excitation energies shows how the exciton-phonon interaction behaves, unveiling the scattering strength. The origin of the interaction is in the condition of resonance, which is directly observed in the temperature resolved spectra. Most importantly, the underlying graphene strongly enhances the coupling of phonons and excitons. Consequently, an enhanced resonant Raman spectrum of GaN nanocrystals possessing clearly observable phonon overtones up to the $4^{th}$ order has been obtained. It has been demonstrated that the responsible effect is the electron transfer between nanocrystals and the underlying graphene. The utilization of such an increased coupling effect can be beneficial for a study of the charge carrier scattering in semiconducting nanomaterials, analysis of their crystal quality, improvement of sensor sensitivity and in the subsequent development of new-generation optoelectronic devices.*

**TEXT**

The electron-phonon interaction, one of the most widely explored topics in solid state physics, has been extensively studied due to its influence on the basic physical properties of solids, including electron mobility, electrical resistance or charge carrier relaxation. This phenomenon is emphasized in polar semiconductors where excited electrons, which behave as excitons even at room temperature[1], strongly interact with longitudinal lattice vibrations, giving rise to the exciton-phonon interaction (EPI). The so called Fröhlich mechanism of exciton-LO phonon coupling is a dominant factor of the EPI in III-V semiconductors. This coupling modifies the binding energy of the excitons, which is reflected in the photoluminescence (PL) spectrum, where the energy shift caused by LO phonons is observable in the form of phonon overtones accompanying the broad PL band. While this behaviour has been extensively investigated for a plethora of materials, its modification by graphene is still poorly understood.

With the trend towards the miniaturisation of electronic devices, nanomaterials offer novel electronic and optical properties that differ from the well-studied bulk materials. The EPI has been observed in a broad range of nanomaterials made of CdS, ZnO, $MoSe_2$, ZnTe and several other polar semiconductors[2–4]. Its strength and characteristics varies not only among these materials, but also with respect to the dimension and size of the nanostructures[2,5,6]. It can be further enhanced by various mechanisms, like in multivalley materials[7], interfaces[8,9] or in heterostructures, where local electric fields affect the EPI[10]. The tuning of EPI may be superior in the research of superconductors or integrated optoelectronics.

Gallium nitride (GaN) is a semiconductor with a direct bandgap of 3.4 eV[11], commonly used for UV light emission. Due to a mismatch of standard substrates and the wurtzite lattice of GaN, there are efforts to grow monocrystalline GaN on novel substrates[12]. Graphene is a suitable candidate, as the lattice of this two-dimensional carbon-based material is also hexagonal. The attempts to grow GaN on graphene have led to a formation of nanoislands[13], nanocrystallites[13,14], nanocrystals[15,16] or nanowires[17,18] with high crystallinity but a specific dimensionality. The structural quality of the produced GaN nanostructures can be studied using Raman spectroscopy, uncovering the character of bulk and surface phonons under both non-resonant and resonant conditions[14,16,17]. In a specific scenario when the excitation energy exceeds the bandgap energy, the resonance condition is fulfilled and the resonant Raman scattering is observed. Concerning the fact that GaN shows a nonnegligible EPI[19], the Fröhlich interaction (one of the contributions to EPI) takes place, leading to a distinct set of phonon overtones accompanied by PL peak in resonant Raman scattering spectrum[14,20–22]. Such a behaviour has been described in several GaN-based nanostructures with the phonon overtones up to the 4th order, indicating on a superior crystalline quality[18,19]. However, despite this high quality, the intensities of the 3rd and 4th overtones were low, with the last one hardly detectable.

In literature, the utilization of graphene substrate for an increase of the Raman signal out of small nanostructures has been widely reported. In this technique, known as graphene enhanced Raman spectroscopy (GERS), the transfer of electrons from graphene to nanostructures is responsible for the signal enhancement[23–26]. However, until now, in these studies almost exclusively non-resonant Raman spectroscopy has been applied and, especially, no work on enhancement of the phonon overtones by graphene has been reported.

In this paper, we report on utilization of graphene substrates for an enhancement of resonant Raman spectra of GaN nanocrystals. Although the occurrence of EPI in GaN (without graphene substrate) has been thoroughly described[14–16], the influence of the underlying graphene on the scattering mechanism is still poorly understood. Therefore, we have performed experiments to show how graphene affects the EPI in GaN nanocrystals, and we address the origin of the enhancement of the Raman and PL signal as an effect of the transfer of the electrons from graphene to GaN nanocrystals.

The fabrication procedure started with CVD graphene transfer onto an isolating $SiO_2$ (280 nm) layer on a Si(100) substrate by using the PMMA-assisted wet transfer. As the $SiO_2$ substrate was only partially covered by graphene, it made it possible to grow GaN nanocrystals on two different substrates, graphene/$SiO_2$ and bare $SiO_2$ only. The nanocrystals were grown using droplet epitaxy[27] in two consecutive steps. First, Ga droplets were deposited by an effusion cell Omicron EMF 3 from a PBN crucible, while the sample was heated to 200 °C to increase the surface diffusion length. Afterwards, Ga droplets were nitridized by hyperthermal nitrogen ions (E ≤ 50 eV) at low substrate temperature of 200 °C using an ion-atomic beam source designed by our group[55]. The nitrogen ion flux supplied a current density of 1 uA/cm$^2$ at a partial pressure of $1\times10^{-4}$ Pa. The fabricated nanostructures are shown in Fig. 1a as a scanning electron microscopy (SEM) image and in Fig. 1b as an atomic force microscopy (AFM) topography, with a vertical profile of the nanocrystal shown in Fig. 1c. The images were obtained with the SEM Verios 360L from FEI and the AFM Bruker Dimension FastScan. The nanocrystal size and surface density can be controlled by substrate temperature and deposition time[28], in our experiments set to 200 °C and 2h, respectively. The surface density also depends on the diffusion length at the substrate surface, which is smaller for CVD graphene compared to Si(111) [29], resulting in a densely covered surface (coverage 50 %). The substantial surface coverage is necessary for a sufficiently high light-nanocrystals interaction to enable the detection of EPI.

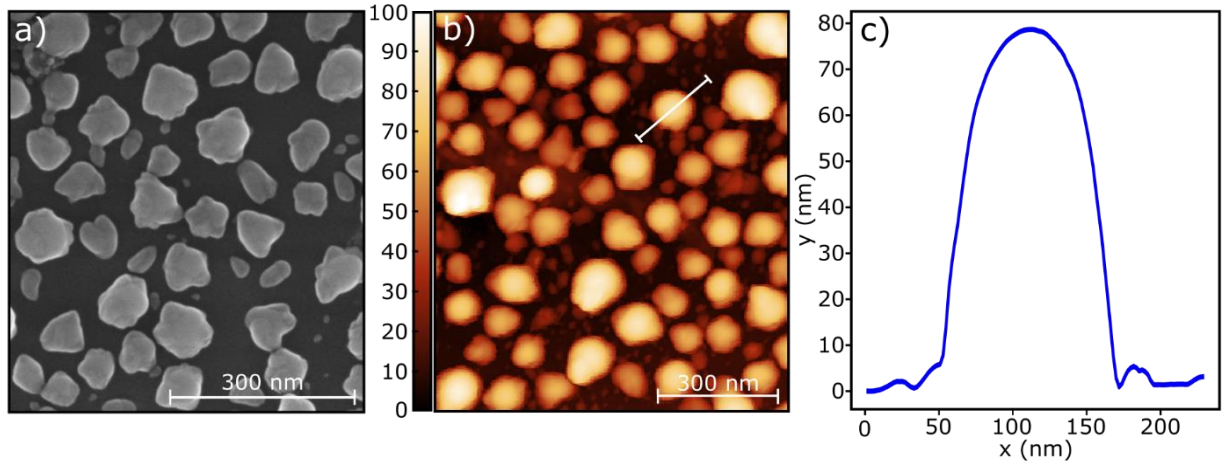

*Fig. 1: a) High resolution image of the GaN nanocrystals from SEM, showing the crystalline nature of the nanostructure, b) Topography of the nanocrystals obtained by AFM measurement and c) vertical profile of the nanocrystal extracted from the AFM image along the white line segment.*

The optical properties of the nanocrystals were measured using a confocal Raman spectrometer (Witec Alpha 300R) with a 100x objective and equipped with 532 nm and 355 nm lasers operated at an optical power of 10 mW and 5 mW, respectively. For the application of additional excitation energies, another Raman spectrometer (NT-MDT NTegra Spectra) was used having He-Cd lasers of 325 nm and 442 nm wavelengths, operated at 10 mW optical power.

GaN forms a hexagonal wurtzite structure described by a space group $C_{6v}^4$ with four atoms in a unit cell, leading to 12 phonon modes. Four of them are Raman active[30]: non-polar modes $E_2^{\text{low}}$ and $E_2^{\text{high}}$ and polar modes $A_1$ and $E_1$. The polar modes create a macroscopic polarization field due to the respective movement of the Ga and N sublattices (unlike the non-polar modes), and thus are split into LO and TO modes. Here only the most prominent $A_1$(LO) mode will be studied in detail. As the Fröhlich mechanism of EPI only enhances LO phonon modes, the mechanism of the increased scattering is confirmed by the enhancement of the $A_1$(LO) phonon and the absence of the $A_1$(TO) phonon in the resonant Raman spectra.

In Fig. 2a the comparison of the Raman spectra of GaN nanocrystals on graphene measured at multiple excitation energies is shown. The key factor is the width of the bandgap of GaN being 3.4 eV, separating these spectra to below-bandgap ($E_{exc} < E_g$) and above-bandgap ($E_{exc} > E_g$) excitations, resulting in non-resonant and resonant Raman scattering. As for the non-resonant scattering, the spectra include the first order Raman peaks of the Si substrate ($\omega$=525 cm$^{-1}$ and $\omega$=990 cm$^{-1}$), graphene peak D ($\omega$=1350 cm$^{-1}$), G ($\omega$=1590 cm$^{-1}$) and 2D ($\omega$=2700 cm$^{-1}$) and a weak $A_1$(LO) peak of GaN ($\omega$=740 cm$^{-1}$). The graphene peaks are distorted due to the fabrication process, where the graphene layer is bombarded by N ions generating lattice defects[31–33]. The presence of $A_1$(LO) peaks indicates a crystalline nature of the nanocrystals, confirming the completeness of the nitridation process. When the condition of resonance is fulfilled, the electron excitation and thus absorption of excitation radiation in GaN significantly increases[34]. Consequently, the scattering at the substrate (graphene, Si), is almost suppressed due to the high surface coverage of the absorbing GaN nanocrystals. Most importantly, the $A_1$(LO) peak is enhanced, as the light is scattered with the participation of electrons excited to the conduction band of GaN. These then interact with LO phonons via the Fröhlich mechanism of EPI. This is enabled in the above-bandgap excitations because of the increased density of excited electrons. Alongside the $A_1$(LO) peak enhancement, also the multi-phonon processes are favourable. They are present in the spectra as phonon replicas $nA_1$(LO) with the Raman shifts $\omega_1$=735 cm$^{-1}$, $\omega_2$=1462 cm$^{-1}$, $\omega_3$=2191 cm$^{-1}$ and $\omega_4$=2916 cm$^{-1}$. The origin of the multi-phonon processes is explained by a cascade model[35], describing the electron energy transitions upon light illumination, as shown in Fig 2b. In this process, electrons in the valence band are lifted to the conduction band upon gaining energy from the photons. This is illustrated as a vertical transition from *a* to *b*. As the excited electrons tend to relax to the low energy edge of the conduction band, they undergo a series of non-radiative transitions with the creation of LO phonons, illustrated as transitions *b* to *c*. After reaching the conduction band edge, electrons relax to the valence band with an emission of PL photons, labelled as a transition *c* to *d*. Therefore, the intensities of given phonon peaks in the obtained Raman spectrum indicate on the abundance of the LO phonons in the relaxation process of photoexcited electrons. An interesting consequence of this phenomenon is that non-zero wavevector phonons participate in the scattering process, as shown in Supplementary Materials.

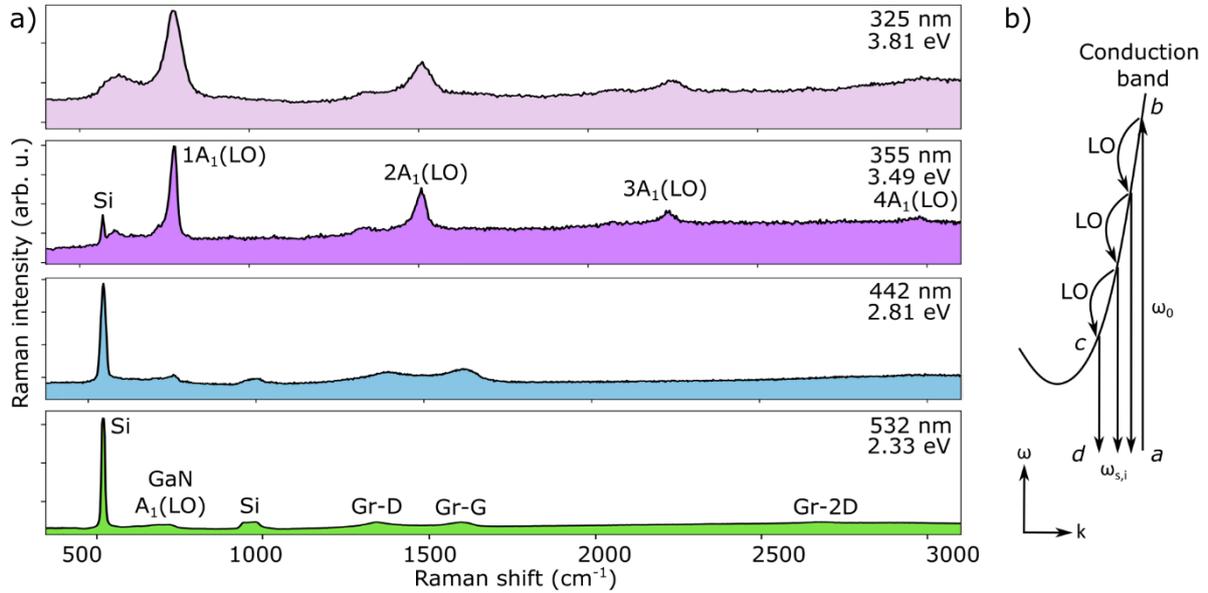

*Fig 2: a) Comparison of Raman spectra of GaN/graphene on the SiO$_2$/Si(100) measured at multiple excitation energies, showing how the resonance governs the interaction of phonons and electrons, b) schematic illustration of the Cascade model.*

To further characterize the interaction of excited electrons with phonons, GaN nanocrystals were studied by temperature resolved Raman spectroscopy at elevated temperatures of up to 300 °C. The obtained spectra including a PL band and phonon replicas are shown in Fig. 3a. The UV excitation laser (λ = 325 nm) was utilized to see the separation between the phonon peaks and the PL band. Extracted energies of the PL peaks (fitted using the Voigt profile) displayed in Fig. 3b demonstrate that the bandgap width of GaN nanocrystals decreases with temperature, in agreement with the previous studies[36,37]. In the simplest empirical approach[38], the bandgap energy is a function of temperature as $E_g(T) = E_{g,0} - aT$ (for $T \gg 0$), where $E_{g,0}$ is the bandgap energy estimated at 0 K and $a$ defines the slope of the linear function. As for this work, the slope is $a$ = 0.031 nm/K (≈ 2.6×10$^{-4}$ eV/K), which corresponds to the experimental values known for other related semiconductors[37,39]. Notably, the PL energy does not correspond with the bandgap of bulk GaN. Although the origin of PL in the GaN nanocrystals is not fully explained, two mechanisms are proposed: the lowering of the PL energy is due to the dopant levels in the bandgap and due to the low dimensionality of the GaN nanocrystals, which modifies the optical response.

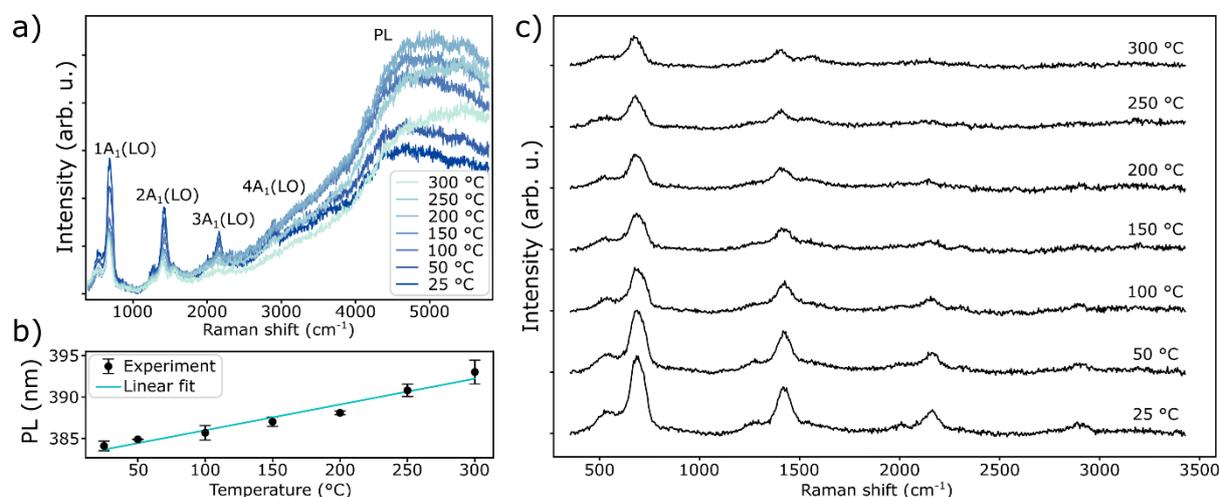

*Fig. 3: a) Temperature resolved Raman spectra, showing phonon replicas of the $A_1(LO)$ phonon along with the PL band, b) the narrowing of the bandgap with temperature, c) isolated Raman replicas of the $A_1(LO)$ peak with respect to temperature, supporting the idea of the EPI.*

Finally, the phonon peaks were isolated from the spectrum by subtracting the broad PL band. The evolution of the phonon peak series with temperature is shown in Fig. 3c. With the increasing temperature, the higher order phonon replicas gradually disappear. It is in contrast with thermal properties of phonons, where the phonon concentration should increase with temperature, as the population of phonons is governed by the Bose-Einstein distribution. Therefore, the origin of the phonon peak diminution can be related to the already described low-energy shift of the PL maxima. The relation between the excitation energy and the bandgap width indicates that the excitation energy must be in a proximity of the bandgap energy in order to observe a pronounced EPI. As the semiconductor is heated up, the bandgap energy decreases and the PL maximum shifts away from the phonon replicas, resulting in a weak coupling between the two excitations. Such a mechanism of a modification of the EPI strength by temperature has been published by Iqbal in ZnTe thin films[40]. Additionally, the thermal extinction of excitons should play only a minor role in this observation, since the evolution of the width of the phonon replicas with temperature is not significant. The thermal properties of phonons are discussed in Supplementary Materials (see Fig. S2).

So far, the findings were related to the structure of GaN nanocrystals on the graphene substrate. To see the influence of graphene on EPI, two spectra were obtained from the same sample with equally sized nanocrystals but only partially covered by graphene. Two arrangements emerged from that, GaN nanocrystals deposited on graphene/SiO$_2$ and directly on SiO$_2$. The comparison of their Raman spectra, utilizing an UV excitation laser of λ = 355 nm, is shown in Fig. 4a. First, graphene induces an enhancement of the PL background of GaN nanocrystals. Such an observation cannot be explained by plasmonic enhancement due to graphene plasmons, as previously published elsewhere[44,45]. The reason is that the width of GaN nanocrystals is > 50 nm (see Fig. 1), and thus it can be hardly supposed that a surface corrugation of a lateral periodicity of ~1 nm needed for generation of UV surface plasmon polaritons in graphene can be achieved. Thus, the PL enhancement is most probably caused by a transfer of electrons from graphene into GaN nanocrystals, as will be discussed within the following paragraph. Nevertheless, for a precise understanding of the mechanism, further detailed study will be necessary. This idea is supported by the fact that the work function of GaN is typically larger than that of the monolayer graphene and thus the electrons should move from graphene to GaN nanocrystals [46]. More importantly, alongside the PL background, the series of phonon overtones was enhanced. Notably, not only the increased intensity of the first and second overtone (735 and 1462 cm$^{-1}$, resp.), but even higher order overtones were observed, located at 2191 and 2916 cm$^{-1}$.

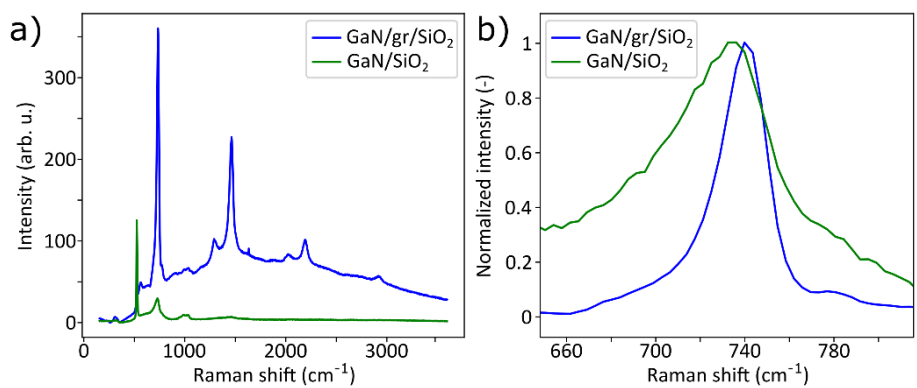

***Fig. 4:*** *a) Comparison of the Raman spectra of GaN nanocrystals on graphene and on SiO$_2$ obtained for the 355 nm excitation laser, showing how the presence of graphene enhances the EPI, b) The blue-shift of the normalized A$_1$(LO) phonon peak in the presence of graphene is given by the LOPC, which indicates on a charge carrier density change[47].*

The influence of graphene on the increased exciton-LO phonon scattering can be once more qualitatively explained by an enhanced electron density in GaN nanocrystals. This quantity is of a major importance in the scattering rate, which in turn rules the transition probability in the cascade model. Evidence of the enhanced charge carrier density is based on the observation shown in Fig. 4b. Here the $A_1(LO)$ peak from both spectra (on graphene and $SiO_2$) is compared, showing an observable blue-shift of 5 cm$^{-1}$ with the presence of graphene. The shift is caused by a strong coupling between longitudinal phonons and longitudinal plasmons ($\varepsilon = 0$) in a spatial frequency range of hundreds of cm$^{-1}$ leading to new hybrid modes (LOPC – longitudinal optical phonon-plasmon coupling) represented by the upper $L^+$ and a lower $L^-$ branch in the dispersion relation similarly to that shown by Harima[47]. The branches are split from each other (anticrossing) and the value of this so-called Rabi splitting is a measure of the coupling strength[48]. For lower charge carrier densities (i.e. lower plasma frequency), the upper $L^+$ branch modes are identical with LO phonons (i.e. $A_1(LO)$ in our case). However, as the densities increase, the $L^+$ branch modes start to deviate from the uncoupled LO phonons towards higher energies. Hence, in case of doping provided by graphene underneath of GaN nanocrystals, one can observe the blue-shift depicted in Fig. 4b. Such a behaviour has been already experimentally described for doped GaN as well as other III-V semiconductors[47,49–52].

The charge carrier (exciton) density in GaN nanocrystals on graphene is defined by two factors: the conditions during the deposition and the already mentioned electron transfer over the interface of the nanocrystals and graphene. As for the deposition conditions, the concentration of dopants and impurities in the growth process is critical, while the latter factor reflects the fact that graphene inevitably induces doping due to its electronic properties of the zero-gap material. More specifically, the interface of GaN and graphene can be modelled by a Schottky junction, through which an electron transfer occurs. This mechanism is based on a transfer of charges between the graphene Fermi level and the energy levels of the system located at the graphene surface[23–26,53,54].

This paper has given a thorough description of the exciton-phonon interaction in GaN nanocrystals. By comparing the Raman spectra obtained with varying excitation energies, it was shown that the nature of the light scattering enhancement is governed by the interaction of excitons with lattice vibrations. The resonant character of the interaction is further captured in the temperature resolved Raman spectra, where the thermal shift of the bandgap is observed to affect the coupling strength. Yet most importantly, the influence of a graphene substrate on the above-described interaction has been presented. Using such a substrate, enhanced resonant Raman spectra of GaN nanocrystals possessing clearly observable phonon overtones up to the 4$^{th}$ order have been obtained. It has been demonstrated that the responsible effect is the electron transfer between nanocrystals and the underlying graphene. The experimental evidence for an increase of the charge carrier density when graphene is used as the substrate for GaN nanocrystals is given both by an enhanced photoluminescence (see Fig. 4a) and by the increase of the spatial frequency of the $A_1$(LO) phonons (see Fig. 4b). Such a shift is caused by a charge carrier (electron) density increase in GaN and consequently by the increased LOPC giving rise to new hybrid phonon-plasmon modes. Better understanding and mastering of the graphene enhanced resonant Raman scattering represents a valuable step towards advanced applications of GaN nanocrystals and other semiconductor nanostructures in optoelectronics and photonics.

## SUPPLEMENTARY MATERIAL

In Supplementary Material, the non-zero wavevector phonons are discussed to support the cascade model. Additionally, the thermal properties of phonons are studied to show the phonon softening.

## ACKNOWLEDGEMENT

We acknowledge the support of Czech Science Foundation (Grant No. 23-07617S) and of CzechNanoLab Research Infrastructure supported by MEYS CR (LM2023051). This work was also supported by the project Quantum materials for applications in sustainable technologies (QM4ST), funded as project No. CZ.02.01.01/00/22_008/0004572 by OP JAK, call Excellent Research.

## CONFLICT OF INTEREST

The authors have no conflicts to disclose.

## AUTHOR CONTRIBUTIONS

M.Kos. performed the fabrication and sample analysis and wrote the manuscript, J.M. conceptualized the study and supervised the research project, M.B. contributed to the methodology of the work, V.M., J.P. and L.S. supported the fabrication. M.Kon. supported the temperature experiment, D.N. contributed with theoretical calculations and T.Š. directed the entire study and revised the manuscript.

**REFERENCES**

1.  Viswanath, A. K., Lee, J. I., Kim, D., Lee, C. R. & Leem, J. Y. Exciton-phonon interactions, exciton binding energy, and their importance in the realization of room-temperature semiconductor lasers based on GaN. *Phys Rev B* **58**, 16333–16339 (1998).

2.  Zhang, Q. *et al.* Exciton-phonon coupling in individual ZnTe nanorods studied by resonant Raman spectroscopy. *Phys Rev B Condens Matter Mater Phys* **85**, (2012).

3.  Qin, L. *et al.* Multiphonon Raman Scattering and Strong Electron-Phonon Coupling in 2D Ternary Cu2MoS4Nanoflakes. *Journal of Physical Chemistry Letters* **11**, 8483–8489 (2020).

4.  Marabotti, P. *et al.* Electron-phonon coupling and vibrational properties of size-selected linear carbon chains by resonance Raman scattering. *Nat Commun* **13**, (2022).

5.  Chang, R. & Lin, S. Field and size dependence of exciton–LO-phonon interaction in a semiconductor quantum dot. *Phys Rev B Condens Matter Mater Phys* **68**, (2003).

6.  Hsu, W. T., Lin, K. F. & Hsieh, W. F. Reducing exciton-longitudinal-optical phonon interaction with shrinking ZnO quantum dots. *Appl Phys Lett* **91**, (2007).

7.  Sohier, T. *et al.* Enhanced Electron-Phonon Interaction in Multivalley Materials. *Phys Rev X* **9**, (2019).

8.  Mahdouani, M. Investigation of the electron-surface phonon interaction effects in graphene on a substrate made of polar materials. *Physica E Low Dimens Syst Nanostruct* **87**, 192–198 (2017).

9.  Choi, I. H. *et al.* Giant Enhancement of Electron–Phonon Coupling in Dimensionality-Controlled SrRuO3 Heterostructures. *Advanced Science* **10**, (2023).

10. Li, Z. *et al.* Boosting Enhancement of the Electron–Phonon Coupling in Mixed Dimensional CdS/Graphene van der Waals Heterojunction. *Adv Mater Interfaces* **9**, (2022).

11. Denbaars, S. P. *et al.* Development of gallium-nitride-based light-emitting diodes (LEDs) and laser diodes for energy-efficient lighting and displays. *Acta Mater* **61**, 945–951 (2013).

12. Kucharski, R., Sochacki, T., Lucznik, B. & Bockowski, M. Growth of bulk GaN crystals. *Journal of Applied Physics* vol. 128 Preprint at https://doi.org/10.1063/5.0009900 (2020).

13. Goswami, L., Pandey, R. & Gupta, G. Epitaxial growth of GaN nanostructure by PA-MBE for UV detection application. *Appl Surf Sci* **449**, 186–192 (2018).

14. Chen, X. B., Morrison, J. L., Huso, J., Bergman, L. & Purdy, A. P. Ultraviolet Raman scattering of GaN nanocrystallites: Intrinsic versus collective phenomena. *J Appl Phys* **97**, (2005).

15. Bessolov, V. N., Zhilyaev, Y. V, Konenkova, E. V, Fedirko, V. A. & Zahn, D. R. T. *Raman and Infrared Spectroscopy of GaN Nanocrystals Grown by Chloride-Hydride Vapor-Phase Epitaxy on Oxidized Silicon. Translated from Fizika i Tekhnika Poluprovodnikov* vol. 37 (2003).

16. Li, H. D. *et al.* Raman spectroscopy of nanocrystalline GaN synthesized by arc plasma. *J Appl Phys* **91**, 4562–4567 (2002).

17. Livneh, T., Zhang, J., Cheng, G. & Moskovits, M. Polarized Raman scattering from single GaN nanowires. *Phys Rev B Condens Matter Mater Phys* **74**, (2006).

18. Dhara, S. *et al.* Multiphonon Raman scattering in GaN nanowires. *Appl Phys Lett* **90**, (2007).

19. Behr, D., Wagner, J., Schneider, J., Amano, H. & Akasaki, I. Resonant Raman scattering in hexagonal GaN. *Appl Phys Lett* 2404 (1995) doi:10.1063/1.116148.

20. Xu, S. J. *et al.* Spectral features of LO phonon sidebands in luminescence of free excitons in GaN. *Journal of Chemical Physics* **122**, (2005).

21. Kaschner, A., Hoffmann, A. & Thomsen, C. Resonant Raman scattering on free and bound excitons in GaN. *Phys Rev B Condens Matter Mater Phys* **64**, (2001).

22. Kaschner, A. *et al.* Evidence for Phase Separation in InGaN by Resonant Raman Scattering. *physica status solidi (a)* **179**, R4–R6 (2000).

23. Naumenko, D., Snitka, V., Snopok, B., Arpiainen, S. & Lipsanen, H. Graphene-enhanced Raman imaging of TiO 2 nanoparticles. *Nanotechnology* **23**, (2012).

24. Xu, H. *et al.* Modulating the charge-transfer enhancement in GERS using an electrical field under vacuum and an n/p-doping atmosphere. *Small* **7**, 2945–2952 (2011).

25. Feng, S. *et al.* Ultrasensitive molecular sensor using N-doped graphene through enhanced Raman scattering. *Sci Adv* **2**, (2016).

26. Persson, B. N. J., Zhao, K. & Zhang, Z. Chemical contribution to surface-enhanced raman scattering. *Phys Rev Lett* **96**, (2006).

27. Mach, J. *et al.* Low temperature selective growth of GaN single crystals on pre-patterned Si substrates. *Appl Surf Sci* **497**, (2019).

28. Kolíbal, M., Čechal, T., Brandejsová, E., Čechal, J. & Šikola, T. Self-limiting cyclic growth of gallium droplets on Si(111). *Nanotechnology* **19**, (2008).

29. Mach, J. *et al.* Electronic transport properties of graphene doped by gallium. *Nanotechnology* **28**, (2017).

30. Azuhata, T., Sota, T., Suzuki, K. & Nakamura, S. Polarized Raman spectra in GaN. *Journal of Physics: Condensed Matter* **7**, L129–L133 (1995).

31. Jorio, A. *et al.* Raman study of ion-induced defects in N-layer graphene. *Journal of Physics Condensed Matter* **22**, (2010).

32. Lucchese, M. M. *et al.* Quantifying ion-induced defects and Raman relaxation length in graphene. *Carbon N Y* **48**, 1592–1597 (2010).

33. Ahlberg, P. *et al.* Defect formation in graphene during low-energy ion bombardment. *APL Mater* **4**, (2016).

34. Muth, J. F. *et al.* Absorption coefficient, energy gap, exciton binding energy, and recombination lifetime of GaN obtained from transmission measurements. *Appl Phys Lett* **71**, 2572–2574 (1997).

35. Martin, R. M. & Varma, C. M. Cascade Theory of Inelastic Scattering of Light. *Phys Rev Lett* **26**, 1241–1244 (1971).

36. Sarkar, N. & Ghosh, S. Temperature dependent band gap shrinkage in GaN: Role of electron-phonon interaction. *Solid State Commun* **149**, 1288–1291 (2009).

37. Zhang, Y., Wang, Z., Xi, J. & Yang, J. Temperature-dependent band gaps in several semiconductors: From the role of electron-phonon renormalization. *Journal of Physics Condensed Matter* **32**, (2020).

38. Varshni, Y. P. *Temperature Dependence of the Energy Gap in Semiconductors*. *Physica* vol. 34 (1967).

39. Cho, Y. *et al.* Temperature dependence on bandgap of semiconductor photocatalysts. *Journal of Chemical Physics* **152**, (2020).

40. Iqbal, M. F., Ain, Q. U., Yaqoob, M. M., Zhu, P. & Wang, D. Temperature dependence of exciton–phonon coupling and phonon anharmonicity in ZnTe thin films. *Journal of Raman Spectroscopy* **53**, 1265–1274 (2022).

41. Tao, Z., Song, Y. & Xu, Z. Probing the thermally driven response of Raman-active phonon modes in sapphire single crystals by in situ Raman spectroscopy. *Ceram Int* **49**, 33175–33187 (2023).

42. Li, W. S., Shen, Z. X., Feng, Z. C. & Chua, S. J. Temperature dependence of Raman scattering in hexagonal gallium nitride films. *J Appl Phys* **87**, 3332–3337 (2000).

43. Xue, X. Y. *et al.* Temperature dependences of Raman scattering in different types of GaN epilayers. *Chinese Physics B* **21**, (2012).

44. Hwang, S. W. *et al.* Plasmon-enhanced ultraviolet photoluminescence from hybrid structures of graphene/ZnO films. *Phys Rev Lett* **105**, (2010).

45. Dong, Z., Chen, P., Li, S., Jiang, Z. & Xu, F. Plasmonic effect for photoluminescence enhancement in graphene/Au/ZnO hybrid structures: Dependence on the surface lateral period of the Au layer. *Mater Res Express* **8**, (2021).

46. Deng, Z. & Wang, X. Strain engineering on the electronic states of two-dimensional GaN/graphene heterostructure. *RSC Adv* **9**, 26024–26029 (2019).

47. Harima, H., Sakashita, H. & Nakashima, S. Raman microprobe measurement of under-damped LO-phonon-plasmon coupled mode in n-type GaN. *Materials Science Forum* **264–268**, 1363–1366 (1998).

48. Břínek, L. *et al.* Plasmon Resonances of Mid-IR Antennas on Absorbing Substrate: Optimization of Localized Plasmon-Enhanced Absorption upon Strong Coupling Effect. *ACS Photonics* **5**, 4378–4385 (2018).

49. Artú, L., Cuscó, R., Ibá, J., Blanco, N. & Gonzá Lez-Dí, G. *Raman Scattering by LO Phonon-Plasmon Coupled Modes in n-Type InP*.

50. Eric N'Dohi, A. J. *et al.* Micro-Raman characterization of homo-epitaxial n doped GaN layers for vertical device applications. *AIP Adv* **12**, (2022).

51. Talwar, D. N. Direct evidence of LO phonon-plasmon coupled modes in n -GaN. *Appl Phys Lett* **97**, (2010).

52. Wieser, N., Klose, M., Dassow, R., Scholz, F. & Off, J. Raman studies of longitudinal optical phonon-plasmon coupling in GaN layers. *J Cryst Growth* **189–190**, 661–665 (1998).

53. Ling, X. *et al.* Can graphene be used as a substrate for Raman enhancement? *Nano Lett* **10**, 553–561 (2010).

54. Zhang, N., Tong, L. & Zhang, J. Graphene-based enhanced raman scattering toward analytical applications. *Chemistry of Materials* **28**, 6426–6435 (2016).

55. Mach, J. *et al.* An ultra-low energy (30-200eV) ion-atomic beam source for ion-beam-assisted deposition in ultrahigh vacuum. *Review of Scientific Instruments* **82**, (2011).